\documentclass[preprint,nofootinbib]{revtex4}%
\usepackage{amssymb}
\usepackage{amsfonts}
\usepackage{amsmath}
\usepackage{amsmath}
\usepackage{graphicx}
\usepackage[usenames]{color}%
\providecommand{\U}[1]{\protect\rule{.1in}{.1in}}
\providecommand{\U}[1]{\protect\rule{.1in}{.1in}}
\definecolor{blue}{rgb}{0,0,1}

\definecolor{red}{rgb}{1,0,0}

\usepackage{hyperref}
\hypersetup{colorlinks=true,allcolors=blue,	pdftitle={Quantum backreaction in AdS3}}

\renewcommand{\(}{\left(}
\renewcommand{\)}{\right)}
\renewcommand{\[}{\left[}
\renewcommand{\]}{\right]}

\begin{document}
\title{Quantum backreactions in (A)dS$_3$ massive gravity and logarithmic asymptotic behavior}
\author{M. Chernicoff$^1$, G. Giribet$^2$, J. Moreno$^3$, J. Oliva$^3$, R. Rojas$^3$, C. Ramirez de Arellano Torres$^3$}
\affiliation{$^1$Departamento de F\'{\i}sica, Facultad de Ciencias, Universidad Nacional Aut\'{o}noma de M\'{e}xico,   A.P. 70-542, CDMX 04510, M\'{e}xico.}
\affiliation{$^2$Department of Physics, New York University,  726 Broadway, New York, NY10003, USA.}
\affiliation{$^3$Departamento de F\'{\i}sica, Universidad de Concepci\'on, Casilla, 160-C, Concepci\'on, Chile.}

\begin{abstract}
We study the interplay between higher curvature terms and the backreaction of quantum fluctuations in 3-dimensional massive gravity in asymptotically (Anti-)de Sitter space. We focus on the theory at the special point of the parameter space where the two maximally symmetric vacua coincide. In the case of positive cosmological constant, this corresponds to the partially massless point, at which the classical theory admits de Sitter black holes and exhibits an extra conformal symmetry at linear level. We explicitly find the quantum corrected black hole geometry in the semiclassical approximation  and show that it induces a relaxation of the standard asymptotic conditions. Nonetheless, the new asymptotic behavior is still preserved by an infinite-dimensional algebra, which, in addition to Virasoro, contains logarithmic supertranslations. Finally, we show that all the results we obtain for the quadratic massive gravity theory can be extended to theories including cubic and quartic terms in the curvature.      
\end{abstract}
\maketitle

\section{Introduction}

An efficient strategy to avoid the difficulty of not having a quantum theory of gravity is to resort to the semiclassical approximation, in which one considers recursively, and to a certain order of the perturbation theory, Einstein equations sourced with the expectation value of the energy-momentum tensor of the matter field content; namely
\begin{equation}
G_{\mu\nu}+\lambda g_{\mu\nu}=8\pi G\left\langle T_{\mu\nu}\right\rangle \ ,\label{EOM}
\end{equation}
where $ G_{\mu\nu}=R_{\mu\nu}-\frac 12 Rg_{\mu\nu} $ is the Einstein tensor and $\left\langle T_{\mu\nu}\right\rangle $ is the expectation value of the energy-momentum tensor for a given state, whose computation requires a specific regularization scheme. These equations are usually solved perturbatively, allowing to evaluate the corrections to the gravitational
potential coming from quantum effects of matter. This captures the loop corrections to the graviton dynamics. The natural parameter in the expansion is $\kappa =(8\pi G)^{1/2}$, which  controls both the coupling to matter and the gravitational self-interactions. Consequently, self-consistency demands a sort of large $N$ limit in order to consider only the matter contribution in the loop expansion. 

This method has proven extremely useful in the study of various aspects of quantum field theory on curved spacetime, and especially in the study of quantum effects around black holes. Here, we are interested in applying the same method to analyze the interplay between the quantum fluctuations induced by the right hand side of (\ref{EOM}) and the higher-curvature terms that might correct the gravitational dynamics on the left hand side of that equation. 

Higher-curvature terms in the gravitational effective action typically arise due to quantum corrections, and they naturally appear in other approaches such as string theory. It is thus natural to ask what are the backreaction effects of the expectation value $\left\langle T_{\mu\nu}\right\rangle $ when sourcing a higher-order theory of gravity with modified $G_{\mu \nu}$. The main motivation to investigate this problem comes from holography: While in general relativity with negative cosmological constant the leading and sub-leading terms in the standard AdS asymptotic conditions do not get modified by the backreaction of quantum fluctuations, in theories with higher-curvature modification to the Einstein tensor this may be different and result in a weakened fall off close to infinity. More concretely, in models such as massive gravity in $2+1$ dimensions the asymptotic AdS$_3$ conditions can be modified with respect to the standard Brown-Henneaux asymptotic conditions \cite{BrownHenneaux}. Then, since the weakened boundary conditions typically give rise to extra degrees of freedom in the dual CFT and change its properties \cite{LogGravity}, it is worthwhile asking how the backreaction of $\langle T_{\mu \nu}\rangle$ affects the boundary conditions relative to the standard asymptotia. This is in general a very complicated problem whose analytic solution is not at reach, except in very special cases. 

One way to investigate the interplay between quantum fluctuations and higher-curvature terms analytically, in the presence of black holes, and in asymptotically (A)dS space is to consider a ($2+1$)-dimensional toy-model: We will consider the so-called New Massive Gravity (NMG) \cite{NMG, NMG2}, which provides an excellent setup for asking questions such as: {\it a}) how do semiclassical effects from the right-hand side of (\ref{EOM}) deform the black hole solutions in a theory that involves higher-curvature modifications to Einstein tensor?, {\it i.e.} what is the explicit form of black hole solutions in the presence of both quantum fluctuations and higher order terms in asymptotically (A)dS$_3$ space?; {\it b}) how do the asymptotic conditions in such spaces get modified relative to the Brown-Henneaux asymptotic conditions when the backreaction of the expectation value of the energy-momentum tensor is included in the field equations?, and {\it c}) what is the asymptotic isometry algebra that generates the group of transformations that preserve such modified asymptotia?

NMG is a higher-curvature theory that describes a parity-even, massive deformation of Einstein theory in $2+1$ dimensions. Because of this, the analysis we propose here can also provide some information about how the semiclassical theory (\ref{EOM}) would be modified when a massive graviton is brought to the game. To make the rules of the game clear, let us say that in this paper we will not consider that the higher curvature terms of NMG are those induced by the presence of the expectation value of the energy tensor, but rather we will consider the theory (\ref{EOM}) having modified the left side of it with the specific higher-order terms that come from the NMG action. This will enable us to have control of the higher-curvature terms and the quantum fluctuation separately. This will permit to analyze different regimes of the perturbative expansion, tuning $G$, $N$ and the graviton mass, $m$, as independent parameters. Moreover, considering the NMG specific higher-order deformation has other advantages: unlike other higher-curvature theories in $2+1$ dimensions, NMG admits asymptotically dS$_3$ black hole solutions \cite{NMG2, OTT}. This occurs at the partially massless point, where the graviton mass saturates the Higuchi bound and the theory exhibits an additional symmetry at linearized level \cite{GabadadzeGiribet}. In turn, this is a very nice setup to study dS black holes in relation to dS$_3$/CFT$_2$ correspondence \cite{DetournayGiribet}. Besides, the theory also exhibits interesting features about AdS$_3$ space, such as weakened near boundary behavior relative to the Brown-Henneaux asymptotic conditions \cite{OTT}. This allows to study AdS$_3$/CFT$_2$ correspondence in a more general framework \cite{GOTT}. All these features of NMG -- i.e. $i$) the existence of dS$_3$ black holes, $ii$) the enhanced symmetry at the partially massless point, and $iii$) the relaxed (A)dS$_3$ asymptotia-- are extended to gravity theories with cubic and quartic terms in the curvature; this is something we prove in section VI. The paper is organized as follows: In section II, we will review NMG and the corresponding properties that are relevant to our work. In section III, we will review the computation of the expectation value of the energy-momentum tensor of matter fields both in dS$_3$ and in AdS$_3$. For simplicity, we will consider the case of $N$ conformally coupled scalar fields. In section IV, we will compute the backreaction of quantum fluctuations and their interplay with the higher-curvature terms of NMG. We will display explicit solutions which exhibit deformations with respect to the vacuum black hole solutions. In section V, we will consider a deformation of the standard (A)dS$_3$ asymptotic conditions induced by the backreaction of quantum fluctuations. In section VI, the generalization to the cubic and quartic theory will be investigated. Holographic aspects of these extended theories will be analyzed in a companion paper \cite{Nosotros2}.

Let us mention that the study of semiclassical theory (\ref{EOM}) for black holes in $2+1$ dimensions has been studied in the literature; see for example \cite{MartinezZanelli, Casals1, Casals2, Casals3, Casals4, Emparan2020, Emparan20202, Emparannuevo}. 

\section{3D massive gravity}

NMG theory is parameterized by two
coupling constants, $\lambda$ and $m$, the latter being the mass of the
massive, spin-2 local degree of freedom that the theory exhibits. This massive spin-2 mode fulfills the massive Fierz-Pauli equation at linearized level \cite{NMG}. At full non-linear level, the action is given by%
\begin{equation}\label{La2}
S\left[  g_{\mu\nu}\right]  =\frac{1}{16\pi G}\int d^{3}x\sqrt{-g}\left[  R-2\lambda-\frac
{1}{m^{2}}\left(  R_{\mu\nu}R^{\mu\nu}-\frac{3}{8}R^{2}\right)  \right] \ ,
\end{equation}
where the relative coefficient $-3/8$ in the quadratic terms is essential for the nice properties of the theory, e.g. the decoupling of a ghost-like degree of freedom. The field equations coming from (\ref{La2}) read
\begin{equation}\label{eom}
G_{\mu \nu}+\lambda g_{\mu \nu}-\frac{1}{2m^2}K_{\mu \nu}=0,\\
\end{equation}
where the tensor
\begin{equation}\label{elKmunu}
K_{\mu \nu}=2\nabla^2R_{\mu\nu}-\frac{1}{2}(\nabla_\mu \nabla_\nu R+g_{\mu \nu}\nabla^2 R)-8 R_{\mu \rho}R^{\rho}_{\,\,\nu}+\frac{9}{2}RR_{\mu\nu}+\frac{1}{8}g_{\mu\nu}\left(24R^{\alpha \beta}R_{\alpha \beta}-13R^2\right)
\end{equation} 
is of fourth order in the metric, and it satisfies 
\begin{equation}
K_{\mu \nu}g^{\mu\nu}=K=  R_{\mu\nu}R^{\mu\nu}-\frac{3}{8}R^{2}; \label{latter}
\end{equation}
i.e. its trace turns out to be proportional to the Lagrangian density from which it derives. The latter property is crucial to prove the consistency of the 3D theory \cite{NMG}.

For generic values of the coupling constants, NMG theory admits two maximally symmetric
vaccua with constant curvature $R_{\ \ \lambda\rho}^{\mu\nu}=\Lambda
\delta_{\lambda\rho}^{\mu\nu}$, with $\Lambda $ admitting two values
\begin{equation}
\Lambda_{\pm }=2m^{2}\left(  1\pm\sqrt{1-\frac{\lambda}{m^{2}}}\right)  \ .\label{vacua}
\end{equation}
When $m$ is large and $\lambda $ is fixed, we get
\begin{eqnarray}
\Lambda_- \simeq \lambda + \mathcal{O}(\lambda^2/m^2)\, ,\ \ \ \ \ \ \Lambda_+ \simeq 4m^2-\lambda + \mathcal{O}(\lambda^2/m^2)\, .
\end{eqnarray}
$\Lambda_+$ has to be regarded as a non-perturbative solution, analogous to the self-accelerated solutions in higher-order theories. On the curve of the parameter space defined by $\lambda=m^{2}$, the two vaccua (\ref{vacua}) coincide and the space of solutions
is enlarged. In this case, we have
\begin{equation}
\Lambda = \Lambda_{\pm} =2m^2=2\lambda\, .\label{ElUno}
\end{equation}
In particular, when (\ref{ElUno}) is fulfilled, the theory contains hairy black holes for all values of the effective cosmological constant, including asymptotically dS$_3$ black holes \cite{OTT, NMG2}. Moreover, when $\lambda=m^{2}$ the theory around a
maximally symmetric background exhibits, besides diffeomorphisms, a new gauge symmetry at linearized level; this is given by the conformal transformation \cite{GabadadzeGiribet}
\begin{equation}
\delta g_{\mu\nu}(x)\,=\,\omega(x)\, \gamma_{\mu\nu}\, ,
\end{equation}
where $\gamma_{\mu\nu}$ is the metric on the constant curvature background,
and $\omega(x)$ is a local infinitesimal parameter. When the effective cosmological constant is positive ($\Lambda >0$) the point $\lambda=m^2$ coincides with the Higuchi bound $\Lambda=2m^2$, where the theory becomes partially massless. 

The theory about AdS$_3$ for $\lambda=m^2$ is also quite interesting as it admits a set of black hole solutions of which BTZ is a particular case \cite{BTZ}. BTZ is interesting and simple enough to evaluate the quantum backreaction on it and obtain a close expression. Given the fact that the BTZ black hole is constructed as a quotient
of AdS$_3$ \cite{BTZ2}, one can obtain the relevant Green function
for the conformally coupled scalar field on the black hole spacetime in a closed manner. This is done by imposing
identifications on the Green function for the same operator on AdS$3$, for a given boundary condition; see below. With such expressions at hand,
one can evaluate the $\left\langle T_{\mu\nu}\right\rangle $ on the BTZ black
hole geometry and then compute the semiclassical backreacktion of these
quantum fields. This paper is devoted to the
exploration of such scenario in NMG.

\section{The background and the quantum energy-momentum tensor}

Consider $\Lambda=2m^ 2\equiv -1/l^2<0$. The static, constant-curvature black hole solution in AdS$_3$ is the non-rotating BTZ, whose metric is given by
\begin{equation}\label{back}
ds^2=-\left(-\Lambda r^2-\mu+1\right)dt^2+\frac{dr^2}{\left(-\Lambda r^2-\mu+1\right)}+r^2d\phi^2\ ,
\end{equation}
with $-\infty<t<\infty$, $0\leq r<\infty$, and $0\leq\phi\leq2\pi$. This is of course a solution of NMG; $\mu$ is an integration
constant. For $\mu=0$, the spacetime is maximally symmetric and reduces to
AdS$_3$.

Now, consider a conformally coupled scalar field, $\psi$, propagating on this geometry. This fulfils the field equation%
\begin{equation}
\square\psi-\frac{1}{8}R\psi=0\ , \label{feq}%
\end{equation}
with $R$ being the Ricci scalar of (\ref{back}), namely $R=-6/l^2$. Using the field equation (\ref{feq}), the classical energy-momentum tensor
for such a field on the background (\ref{back}) can be written as
\begin{equation}
T_{\mu\nu}=\frac{3}{4}\nabla_{\mu}\psi\nabla_{\nu}\psi-\frac{1}{4}\psi
\nabla_{\mu}\nabla_{\nu}\psi-\frac{1}{4}g_{\mu\nu}\left[  \nabla_{\alpha}%
\psi\nabla^{\alpha}\psi-\frac{\Lambda}{4}\psi^{2}\right]  \ .
\end{equation}
The quantum energy-momentum tensor $\langle {T}_{\mu\nu}\rangle
$ is similarly given by%
\begin{equation}
\langle {T}_{\mu\nu}\rangle =\frac{\hbar}{4}\lim_{x^{\prime}\rightarrow
x}\left(  3\nabla_{\mu}\nabla_{\nu}^{\prime}-\nabla_{\mu}%
\nabla_{\nu}^{\prime}-g_{\mu\nu}\nabla^{\alpha}\nabla_{\alpha}^{\prime}%
+\frac{g_{\mu\nu}}{4}\Lambda\right)  G\left(  x,x^{\prime}\right)  \ ,\label{vevtmunu0}
\end{equation}
where we have resorted to point-splitting to compute the Green function $G\left(  x,x^{\prime}\right)
$. The Green function of BTZ can be obtained summing over modes of the quantum scalar, nevertheless, it can also be easily determined from the fact that
(\ref{back}) can be obtained as an identification of global AdS$_3$. We work with transparent
boundary conditions at infinity. The maximally symmetric background corresponds to the induced
geometry on the locus of a quadratic surface embedded on flat 4-dimensional space of
signature $\left(-,-,+,+\right)$ with coordinates $X^a$. Then, one essentially uses the $r^{-1}=\left(X_aX^a\right)^{-1/2}$ Green function of the
standard three-dimensional Laplace operator, projected over the
quadratic surface, leading to the Green function of a conformally coupled scalar
field on the maximally symmetric background. Finally, to obtain the
Green function on the geometry (\ref{back}), one sums over the
images induced by the identifications, see \cite{MartinezZanelli} and references thereof. This yields the expression
\begin{equation}
\langle {T}_{\ \nu}^{\mu}\rangle =\frac{\hbar F\left(  M\right)
}{r^{3}}\left(  \delta_{t}^{\mu}\delta_{\nu}^{t}+\delta_{r}^{\mu}\delta_{\nu
}^{r}-2\delta_{\phi}^{\mu}\delta_{\nu}^{\phi}\right)  \ , \label{vevtmunu}
\end{equation}
where we introduced $M\equiv\mu-1$. The specific form of $F\left(  M\right)  $ depends on whether the
background (\ref{back}) represents a black hole ($\mu>1$) or a naked
singularity ($\mu\leq 1$), as the sum over images is different in each case. It is important to notice that on
a maximally symmetric background $\langle {T}_{\mu\nu}\rangle
$ vanishes, which is a direct consequence of the absence of conformal anomalies in $2+1$-dimensions. The function
$F\left(  M\right)  $ is positive and bounded; it has a local maximum at $M\sim0.65$, where $F_\text{max}\sim 0.023$, and it becomes exponentially suppressed at large $M$. Also, we know that $F(M)$ and its derivative are continuous at $M=0$, where $F(0)={\zeta(3)}/({2\pi^3})$; see \cite{Casals4} and references therein and thereof.

\section{Semiclassical massive gravity equations}

The semiclassical NMG equations are given by
\begin{equation}\label{eom}
R_{\mu \nu}-\frac 12 Rg_{\mu \nu }+\lambda g_{\mu \nu}-\frac{1}{2m^2}K_{\mu \nu}=8\pi G N\langle {T}_{\mu \nu}\rangle ,\\
\end{equation}
where $K_{\mu \nu}$ was defined in \eqref{elKmunu} and the quantum energy-momentum tensor $\langle {T}_{\mu \nu}\rangle$ comes from (\ref{vevtmunu0}) and yields (\ref{vevtmunu}). $N$ is the number of scalars, which here will be considered large. Since we are interested in the backreaction of quantum fluctuations of $N$ fields on the static BTZ geometry, we will assume the most general circularly symmetric, static ansatz for the metric, which contains two arbitrary functions and can be parameterized as follows
\begin{equation}
ds^2=\left(1+\Omega(r)\right)\left[ -\left(-\Lambda r^2-\mu+1+f_1(r)\right)dt^2+\frac{dr^2}{\left(-\Lambda r^2-\mu+1+f_1(r)\right)}+r^2d\phi^2\right]\ .
\end{equation}
We expand the field equations keeping only the first order terms in $\Omega$ and $f_1$. That is to say, we expand the field equations around the BTZ solutions, which is locally of constant curvature. As shown in \cite{GabadadzeGiribet}, at the special point $m^2=\lambda$ NMG presents a conformal symmetry when expanded around a constant curvature spacetime. Therefore, we can use such symmetry to fix the gauge freedom by setting $\Omega(r)=0$. If we do so, the solution to the field equation reads
\begin{equation}\label{metrica}
ds^2=-f(r)dt^2+\frac{dr^2}{f(r)}+r^2d\phi^2\ ,
\end{equation}
with
\begin{equation}
f\left(  r\right)  =-\Lambda r^{2}-\mu+1-\frac{2\Lambda Nl_{p}}{\mu-1} F\left(  \mu-1\right)  \left(r \log\left( r\sqrt{|\Lambda|}\right)-r\right)  \ , 
\end{equation}
where we have defined the 3-dimensional Planck length $l_p=8\pi G \hbar$. For the black hole, we have $\mu>1$.

We will be mainly interested in the asymptotically AdS$_3$ case. Writing everything in terms of $l=1/\sqrt{-\Lambda }$ and $M=\mu-1$, the metric function reads
\begin{equation}
f\left(  r\right)  =\frac{r^{2}}{l^2}-M+\frac{2Nl_{p}F\left(M\right)}{Ml^2}   \left(r \log\left(\frac{r}{l}\right)-r\right)  \ .\label{postalin}
\end{equation}
As in 3-dimensional Einstein gravity \cite{MartinezZanelli}, the quantum backreaction of the scalar is exponentially suppressed for large black holes. Notice also that our solution \eqref{metrica} is accurate up to terms of the order $\mathcal{O}(N^2l_{p}^2)$. 

The main result here is (\ref{postalin}), so let us make some comments about it: First, we observe that, in contrast to Einstein gravity, where the backreaction of quantum fluctuations only induces a subleading term in $f$ of order $\sim \mathcal{O}(1/r)$, the quadratic theory exhibits terms with a weaker falloff $\sim \mathcal{O}(r)$ and $\sim \mathcal{O}(r\log r)$, even weaker than the 3-dimensional Newtonian term $\sim \mathcal{O}(1)$. This can be interpreted as an amplification of the quantum fluctuations due to the higher-curvature terms. In the case of conformally coupled matter, a term $\sim \mathcal{O}(1/r)$ in $f$ can easily be understood due to the absence of conformal anomaly in 3 dimensions; it is actually the contribution one expects from the piece $\int^r d\text{Vol}\,T_0^{\,0}\sim 1/r$ in a scale invariant theory. The term of order $\sim \mathcal{O}(r)$ can also be easily understood: such a linear term is also compatible with conformal symmetry, e.g. it appears in the vacuum solution of conformal gravity \cite{ConformalGravity}; besides, it was already known to show up in NMG at the special point $m^2=\lambda$, cf. \cite{NMG2, OTT}. The novelty here was the unexpected contribution $\sim \mathcal{O}(r\log r)$ in (\ref{postalin}). This is even weaker than other logarithmic asymptotic conditions studied in NMG, cf. \cite{AdSwaves, Goya}. 

Let us also notice that the location of the event horizon coming from (\ref{postalin}) is corrected with respect to the solution in Einstein gravity sourced with (\ref{vevtmunu}). In the massive gravity theory the horizon is located at
\begin{equation}
r_+\simeq \sqrt{M}+2\frac{Nl_{p}F\left(M\right)}{2Ml^2}\left(1-\log\sqrt{M}\right)\ .
\end{equation}
For sufficiently large black holes, the horizon size decreases as a product of the quantum effects. This is different from to what happens in Einstein gravity, cf. \cite{MartinezZanelli}.

\section{Asymptotic symmetries}

The asymptotic fall-off conditions that accommodate the quantum backreacted spacetime --at the order discussed above-- are
of the form
\begin{align}
g_{rr}  &  =\frac{l^2}{r^{2}}+\mathcal{O}\left(  \frac{\log r}{r^{3}}\right) \, , \nonumber \\
g_{mn}  &  =r^{2}+\mathcal{O}\left(  r\log r\right) \, , \label{condiciones}\\
g_{rm}  &  =\mathcal{O}\left(  \frac{\log r}{r^{2}}\right)\, .\nonumber
\end{align}

It turns out that this family of spacetimes is preserved by the
action of the Lie derivative along the asymptotic Killing vectors $\eta
=\eta^{\mu}\partial_{\mu}$ of the form%
\begin{equation}
\begin{alignedat}{2} &\eta^+=L^+(x^+)+\frac{l^2}{2r^2}\partial_-^2 L^-(x^-)+\cdots\\ &\eta^-=L^-(x^-)+\frac{l^2}{2r^2}\partial_+^2 L^+(x^+)+\cdots\\ &\eta^r=-\frac{r}{2}(\partial_+ L^++\partial_- L^-)+\cdots \end{alignedat} \label{AKV}%
\end{equation}
Here, $L^{\pm}(x^{\pm })$ are arbitrary functions of $x^{\pm }=t\pm l\phi$. These Killing fields generate two copies of the Witt algebra, which can easily be
checked by expanding $L^{\pm}(x^{\pm })$ in Fourier modes. This shows that the Virasoro symmetry of the boundary dynamics is persistent after the relaxation of the asymptotic conditions. In addition, there are two extra asymptotic Killing vectors preserving the behavior (\ref{condiciones}); these are given by
\begin{equation}
\chi =A\left(  x^{+},x^{-}\right)  \partial_{r}\ \ \ \text{ and }\ \ \ \xi =B\left(  x^{+}%
,x^{-}\right)  \log r\ \partial_{r}\ ,
\end{equation}
with $A(x^{+},x^{-})$ and $B(x^{+},x^{-})$ being two arbitrary functions. While $\chi $ was already observed to appear in \cite{OTT, Donnay}, $\xi $
appears as a new feature due to the quantum back reaction, and hereafter we
will refer to it as the logarithmic supertranslations -- logarithmic supertranslations were already presented in a different context in \cite{FuentealbaHenneaux}. Virasoro algebras generated by $\eta $ are in semidirect sum with that generated by $\chi $ and $\xi$. More specifically, we find the algebraic structure $[\eta ,\chi ]=\chi $, $[\eta ,\xi ]=\xi + \chi $, $[\xi ,\chi ]=0 $. 

\section{Extension to cubic and quartic combinations}

Now, let us show that the features discussed above for NMG at the special point $\lambda=m^2$ persist when one considers cubic and quartic generalizations of it. More precisely, the idea in this section is to show that, whenever the higher-curvature gravity theory is defined on a curve of the parameter space where the different maximally symmetric vacua coincide, then the theory $i$) exhibits an additional conformal symmetry at linearized level, $ii$) admits hairy (A)dS$_3$ black hole solutions, $iii$) exhibits weakened (A)dS$_3$ asymptotia induced by the backreaction of quantum fluctuations. 

Let us start by going back to the NMG quadratic action in order to be systematic
\begin{equation}
	\label{NMG}
	S=\frac{1}{16\pi G}\int_{\mathcal{M}}{d^3x\sqrt{-g}\[R-2\lambda-\frac{1}{m^2}\(R_{\mu\nu}R^{\mu\nu}-\frac{3}{8}R^2\)\]}.
\end{equation}
We expand this action at second order in the perturbation $h_{\mu\nu}$ around a maximally symmetric spacetime given by $R^{\mu\nu}_{\,\, \sigma\rho}=\Lambda(\delta^{\mu}_{\sigma}\delta^{\nu}_{\rho}-\delta^{\mu}_{\rho}\delta^{\nu}_{\sigma})$, with
\begin{equation}
	4m^2\(\lambda-\Lambda\)+\Lambda^2=0\, .
\end{equation}
Notice that so far we are not necessarily considering the special point where the vacuum is unique, but only demanding that the (A)dS$_3$ background satisfies the field equation derived from (\ref{NMG}). The expression for the action $S$ expanded at second order in $h_{\mu\nu}$ is cumbersome and not very illustrative, so we will omit it. We now take the variation of the action under the transformation, expanded at second order in $h_{\mu\nu}$; namely
\begin{equation}
	\delta g_{\mu\nu}=\omega(x^\alpha) \gamma_{\mu\nu} \label{veintitres}
\end{equation}
where $\gamma_{\mu\nu}$ is the metric for the (A)dS$_3$ background (i.e. $g_{\mu\nu}=\gamma_{\mu\nu}+h_{\mu\nu}$, $\delta g_{\mu\nu}=h_{\mu\nu}=\omega\gamma_{\mu\nu}$) and $\omega$ is some arbitrary infinitesimal function. We obtain
\begin{equation}
	\delta S=-\frac{1}{8\pi G}\int_{\mathcal{M}}{d^3x\sqrt{-g}}\[\frac{1}{2} \(1-\frac{\Lambda}{2m^2}\)\(\Box{h}-\nabla_\mu\nabla_\nu h^{\mu\nu}\) +\(\lambda-\frac{\Lambda^2}{4m^2}\)h\]\omega
\end{equation}
which is consistent with \cite{GabadadzeGiribet}. When the vacuum is unique, $\lambda=m^2={\Lambda}/{2}$, and it straightforwardly follows that $\delta S=0$. 

Next, let us see what happens with this strategy when we include $R^3$ and $R^4$ terms in the gravity action. Consider the action 
\begin{equation}
	S=\frac{1}{16\pi G}\int_{\mathcal{M}}{d^3x\sqrt{-g}\[R-2\lambda-\frac{1}{m^2}\(R_{\mu\nu}R^{\mu\nu}-\frac{3}{8}R^2\)+\sigma\(R^\mu_\rho R^\rho_\nu R^\nu_\mu-\frac{9}{8}RR_{\mu\nu}R^{\mu\nu}+\frac{17}{64}R^3\)\]}\nonumber
\end{equation}
which is the natural cubic generalization of NMG \cite{Paulos, Shina, Nosotros2}; $\sigma $ is a coupling constant with mass dimension $-4$. Let us expand at second order in $h_{\mu\nu}$ around a maximally symmetric background. This yields
\begin{equation}
	\Lambda^2+4m^2\(\lambda-\Lambda+\frac{3}{16}\Lambda^3\sigma\)=0  .
\end{equation}
By varying the action under the transformation (\ref{veintitres}), we obtain
\begin{equation}
	\delta S=-\frac{1}{8\pi G}\int_{\mathcal{M}}{d^3x\sqrt{-g}} \[\frac{1}{2}\(1-\frac{\Lambda}{2m^2}-\frac{9}{16}\Lambda^2\sigma\)\(\Box{h}-\nabla_\mu\nabla_\nu h^{\mu\nu}\)+ \(\lambda-\frac{\Lambda^2}{4m^2}-\frac{3}{8}\Lambda^3\sigma\)h\]\omega\nonumber ,
\end{equation}
and when the vacuum is unique, namely for ${m^2}={3\lambda}/{4}$, $\sigma=-{16}/({81\lambda^2})$ and $\Lambda=3\lambda$, we find again $\delta S=0$.

Last, let us proceed in a similar way with the quartic action
\begin{align}\nonumber
	S&=\frac{1}{16\pi G}\int{d^3x}\sqrt{-g}\left[R-2\lambda-\frac{1}{m^2}\(R_{\mu\nu}R^{\mu\nu}-\frac{3}{8}R^2\) + \sigma\(R^\mu_\rho R^\rho_\nu R^\nu_\mu-\frac{9}{8}RR_{\mu\nu}R^{\mu\nu}+\frac{17}{64}R^3\)
 \right. \notag\\ &\quad \ \ \ \ \ \ \ \ \ \ \ \ \ \ \ \ \ \ \ \ \ \ \ \ \  +\left.\nu\( \left(R_{\mu\nu}R^{\mu\nu}\)^2-\frac{4}{3}RR^\mu_\rho R^\rho_\nu R^\nu_\mu +\frac{3}{4}R^2R_{\mu\nu}R^{\mu\nu}-\frac{41}{192}R^4\right)\right],
\end{align}
which, again, is the natural generalization of NMG \cite{Nosotros2}; $\nu$ is a coupling constant of mass dimension $-6$. The variation of the action expanded at second order in the perturbation $h_{\mu\nu}$ around the (A)dS$_3$ background satisfying
\begin{equation}
	\Lambda^2+4m^2\(\lambda-\Lambda+\frac{3}{16}\Lambda^3\sigma-\frac{5}{8}\Lambda^4\nu\)=0
\end{equation}
is
\begin{eqnarray}
	\delta S =&&-\frac{1}{8\pi G}\int_{\mathcal{M}}{d^3x\sqrt{-g}}\Big[\frac{1}{2}\(1-\frac{\Lambda}{2m^2}-\frac{9}{16}\Lambda^2\sigma+\frac{5}{2}\Lambda^3\nu\)\(\Box{h}-\nabla_\mu\nabla_\nu h^{\mu\nu}\)+\nonumber \\
 &&\ \ \ \ \ \ \ \ \ \ \ \ \ \ \ \ \ \ \ \ \ \ \ \ \ \ \ \ \ \ \Big(\lambda-\frac{\Lambda^2}{4m^2}-\frac{3}{8}\Lambda^3\sigma+\frac{15}{8}\Lambda^4\nu\Big)h\Big]\omega .
\end{eqnarray}
Once again, when the vacuum is unique, what happens for $m^2=2\lambda/3$, $\sigma=-1/(3\lambda^3)$, $\nu=-1/(160\lambda^3)$ and $\Lambda=4\lambda$, we find $\delta S=0$.

In other words, this generalizes the result of \cite{GabadadzeGiribet} to the cubic and quartic theory: The linearized theory about (A)dS$_3$, when evaluated on the curve of the parameter space where the maximally symmetric vacua coincide, exhibits a conformal symmetry. Also on that curve of the parameter space, the theory admits hairy black hole with relaxed (A)dS$_3$ asymptotic conditions that are still consistent with Virasoro symmetry at the boundary. It can also be shown that, for theories with arbitrary order in the curvature, whenever the vacua is unique, it is possible to integrate the static solution including the backreaction of the expectation value of the energy-momentum tensor; the results also yields the metric deformation like the one we obtained in (\ref{postalin}); see \cite{Nosotros2} for details of the computation of the $R^n$ theory.

\section{Asymptotic behavior for generic values of the couplings in NMG}

Before concluding, let us elaborate on the quantum backreation in the BTZ background of NMG, but now for generic values of the couplings in NMG. As mentioned above, for generic values ($m^2\neq \lambda $) of the couplings there are two possible BTZ geometries as there are two possible AdS$_3$ radii. Besides, the graviton mass $m$ is not longer related to the bared cosmological constant $\lambda$, so that the quadratic-curvature terms now only introduce short-term modifications. Given the absence of the conformal
symmetry at linearized level, in
order to study the quantum backreaction we have to consider the
following ansatz
\begin{equation}
ds^{2}=-\left(  \frac{r^{2}}{l^{2}}-M+f_{1}\left(  r\right)  \right)
dt^{2}+\frac{dr^{2}}{\frac{r^{2}}{l^{2}}-M+g_{1}\left(  r\right)  }+r^{2}%
d\phi^{2}\ ,
\end{equation}
with $f_1\neq g_1$. In order to obtain compact expressions for these two functions, we solve the bared cosmological
constant $\lambda$ in terms of the AdS$_3$ radii $l$ and the graviton mass $m$, which follows from (\ref{vacua}). The linearized equations for $f_{1}$ and $g_{1}$, sourced by the quantum
renormalized stress-energy tensor, cannot be solved in a closed manner in this case. We
therefore focus on the leading asymptotic term in the $r\rightarrow\infty$
expansion. Assuming the form
\begin{equation}
f_{1}\left(  r\right)  =a_{1}r^{-\delta_{1}}\left(  1+\mathcal{O}\left(
r^{-1}\right)  \right)  \text{ and }g_{1}\left(  r\right)  =b_{1}%
r^{-\delta_{2}}\left(  1+\mathcal{O}\left(  r^{-1}\right)  \right)  \ ,
\end{equation}
in the $tt$ component of the field equations yields
\begin{align*}
&4m^{2}l^{2}r-6Ml^{2}a_{1}(\delta_{1}+2)^{2}r^{-\delta_{1}}+Ml^{2}b_{1}%
(9\delta_{2}^{2}+24+\delta_{2}^{3}+8\delta_{2})r^{-\delta_{2}}\\
&+\delta_{2}%
b_{1}(1+2m^{2}l^{2})r^{2-\delta_{2}}-a_{1}\delta_{1}(\delta_{1}+2)(\delta
_{1}+1)^{2}r^{2-\delta_{1}}=0\ .
\end{align*}
Avoiding any tuning of the couplings, this equation leads to $\delta_{1}=-1=\delta_{2}$. One can check that the large-$r$ behavior of the remaining field equations are fulfilled as
well. Consequently, the large distance behavior of the quantum backreacted
metric reads%
\begin{equation}
ds^{2}=-\left(  \frac{r^{2}}{l^{2}}-M+\frac{l_{p}\alpha_{1}}{r}+\mathcal{O}%
\left(  r^{-2}\right)  \right)  dt^{2}+\frac{dr^{2}}{\frac{r^{2}}{l^{2}%
}-M+\frac{l_{p}\beta_{1}}{r}+\mathcal{O}\left(  r^{-2}\right)  }+r^{2}%
d\phi^{2}\ ,\label{rrrrrr}
\end{equation}
for some given constants $\alpha_{1}$ and $\beta_{1}$. Therefore, for generic
values of the couplings, the quantum backreacted metric in NMG has the same
asymptotic expansion as the one of the quantum corrected BTZ black hole in Einstein gravity, which was of course expected. The reason why our previous solution (\ref{metrica})-(\ref{postalin}) exhibits a logarithmic, slow-decaying correction is that, when evaluating the theory on the special point $\lambda = m^2$, it appears a cooperation between the otherwise short-distance corrections induced by the higher-curvature terms and the large-distance influence of the cosmological constant, ultimately resulting in a large range modification of the asymptotic. For $\lambda \neq m^2$, in contrast, one recovers (\ref{rrrrrr}).

\section{Conclusions}

In this paper we have studied the interplay between higher-curvature terms and the backreaction on the geometry induced by quantum fluctuation. As a working example, we considered 3-dimensional higher-curvature massive gravity theory coupled to $N$ conformally coupled scalar fields in (A)dS$_3$ space. In the large $N$ limit, we solved the semi-classical field equations analytically at first order approximation in $GN$. We have shown that, in contrast to what happens in 3-dimensional Einstein gravity, the backreaction on the higher-curvature theory induces a modification of the (A)dS$_3$ asymptotics relative to the standard Brown-Henneaux boundary conditions. It is worth mentioning that, in the context of holography, this is relevant as relaxation of asymptotic conditions typically lead to extra degrees of freedom in the dual CFT$_2$. For the new set of boundary conditions, we have shown that the asymptotic isometry group is generated by the local conformal algebra in semidirect sum with supertranslations and logarithmic supertranslations. We also extended the analysis to theories containing cubic and quartic terms in the curvature. For such theories, we also obtained the analytic solution of the (A)dS$_3$ black hole including the backreaction of quantum fluctuations, which exhibit the same asymptotic behavior as in the quadratic theory. In a companion work, we study holographic aspects of these theories and possible generalizations \cite{Nosotros2}.


Before concluding, let us mention that quantum corrections in three dimensions were also studied in a different context in \cite{Ferreira:2014ina}. It would be interesting to study the connection to that work in more detail.

\section*{Acknowledgements}

We thank Ana Climent, Nicolás Grandi and Alberto G\"uijosa for their enlightening comments.
The work of M.C. is partially supported by Mexico's National Council of Science and Technology (CONACyT) grant A1-S-22886, and DGAPA-UNAM grant IN116823. The work of J.M. is supported by FONDECYT Postdoctorado Grant 3230626. The research of R.R. is funded by FONDECYT Postdoctorado Grant 3220663. This work was partially funded by FONDECYT Regular Grants 1221504, 1200022, 1200293, 1210500, 1210635.


\begin{thebibliography}{9}

\bibitem{BrownHenneaux}
J.~D.~Brown and M.~Henneaux,
``Central Charges in the Canonical Realization of Asymptotic Symmetries: An Example from Three-Dimensional Gravity,''
Commun. Math. Phys. \textbf{104}, 207-226 (1986).

\bibitem{LogGravity}
A.~Maloney, W.~Song and A.~Strominger,
``Chiral Gravity, Log Gravity and Extremal CFT,''
Phys. Rev. D \textbf{81}, 064007 (2010)
[arXiv:0903.4573 [hep-th]].


\bibitem{NMG}
E.~A.~Bergshoeff, O.~Hohm and P.~K.~Townsend,
``Massive Gravity in Three Dimensions,''
Phys. Rev. Lett. \textbf{102}, 201301 (2009)
[arXiv:0901.1766 [hep-th]].

\bibitem{NMG2}
E.~A.~Bergshoeff, O.~Hohm and P.~K.~Townsend,
``More on Massive 3D Gravity,''
Phys. Rev. D \textbf{79}, 124042 (2009)
[arXiv:0905.1259 [hep-th]].

\bibitem{OTT}
J.~Oliva, D.~Tempo and R.~Troncoso,
``Three-dimensional black holes, gravitational solitons, kinks and wormholes for BHT massive gravity,''
JHEP \textbf{07}, 011 (2009)
[arXiv:0905.1545 [hep-th]].

\bibitem{GabadadzeGiribet}
G.~Gabadadze, G.~Giribet and A.~Iglesias,
``New Massive Gravity on de Sitter Space and Black Holes at the Special Point,''
[arXiv:1212.6279 [hep-th]].

\bibitem{DetournayGiribet}
S.~de Buyl, S.~Detournay, G.~Giribet and G.~S.~Ng,
``Baby de Sitter black holes and dS$_3$/CFT$_2$,''
JHEP \textbf{02}, 020 (2014)
[arXiv:1308.5569 [hep-th]].

\bibitem{GOTT}
G.~Giribet, J.~Oliva, D.~Tempo and R.~Troncoso,
``Microscopic entropy of the three-dimensional rotating black hole of BHT massive gravity,''
Phys. Rev. D \textbf{80}, 124046 (2009)
[arXiv:0909.2564 [hep-th]].

\bibitem{Nosotros2}
M. Chernicoff, G. Giribet, J. Moreno,
J. Oliva, R. Rojas, C. Ramirez de Arellano Torres, 
``Higher-curvature gravity in AdS$_3$, holographic $c$-theorems and black hole microstates,''
to appear.

\bibitem{MartinezZanelli}
C.~Martinez and J.~Zanelli,
``Back reaction of a conformal field on a three-dimensional black hole,''
Phys. Rev. D \textbf{55}, 3642-3646 (1997)
[arXiv:gr-qc/9610050 [gr-qc]].

\bibitem{Casals1}
M.~Casals, A.~Fabbri, C.~Mart\'\i{}nez and J.~Zanelli,
``Quantum dress for a naked singularity,''
Phys. Lett. B \textbf{760}, 244-248 (2016)
[arXiv:1605.06078 [hep-th]].

\bibitem{Casals2}
M.~Casals, A.~Fabbri, C.~Mart\'\i{}nez and J.~Zanelli,
``Quantum Backreaction on Three-Dimensional Black Holes and Naked Singularities,''
Phys. Rev. Lett. \textbf{118}, no.13, 131102 (2017)
[arXiv:1608.05366 [gr-qc]].

\bibitem{Casals3}
M.~Casals, A.~Fabbri, C.~Mart\'\i{}nez and J.~Zanelli,
``Quantum fields as Cosmic Censors in $(2 + 1)$-dimensions,''
Int. J. Mod. Phys. D \textbf{27}, no.11, 1843011 (2018)

\bibitem{Casals4}
M.~Casals, A.~Fabbri, C.~Mart\'\i{}nez and J.~Zanelli,
``Quantum-corrected rotating black holes and naked singularities in ( 2+1 ) dimensions,''
Phys. Rev. D \textbf{99}, no.10, 104023 (2019)
[arXiv:1902.01583 [hep-th]].

\bibitem{Emparan2020}
R.~Emparan, A.~M.~Frassino and B.~Way,
``Quantum BTZ black hole,''
JHEP \textbf{11}, 137 (2020)
[arXiv:2007.15999 [hep-th]].

\bibitem{Emparan20202}
R.~Emparan and M.~Toma\v{s}evi\'c,
``Strong cosmic censorship in the BTZ black hole,''
JHEP \textbf{06}, 038 (2020)
[arXiv:2002.02083 [hep-th]].

\bibitem{Emparannuevo}
R.~Emparan, J.~F.~Pedraza, A.~Svesko, M.~Toma\v{s}evi\'c and M.~R.~Visser,
``Black holes in dS$_{3}$,''
JHEP \textbf{11}, 073 (2022)
[arXiv:2207.03302 [hep-th]].

\bibitem{BTZ}
M.~Ba\~nados, C.~Teitelboim and J.~Zanelli,
``The Black hole in three-dimensional space-time,''
Phys. Rev. Lett. \textbf{69}, 1849-1851 (1992)
[arXiv:hep-th/9204099 [hep-th]].

\bibitem{BTZ2}
M.~Ba\~nados, M.~Henneaux, C.~Teitelboim and J.~Zanelli,
``Geometry of the (2+1) black hole,''
Phys. Rev. D \textbf{48}, 1506-1525 (1993)
[erratum: Phys. Rev. D \textbf{88}, 069902 (2013)]
[arXiv:gr-qc/9302012 [gr-qc]].

\bibitem{ConformalGravity}
J.~Oliva, D.~Tempo and R.~Troncoso,
``Static spherically symmetric solutions for conformal gravity in three dimensions,''
Int. J. Mod. Phys. A \textbf{24}, 1588-1592 (2009)
[arXiv:0905.1510 [hep-th]].


\bibitem{Goya}
G.~Giribet, A.~Goya and M.~Leston,
``Boundary stress tensor and asymptotically AdS3 non-Einstein spaces at the chiral point,''
Phys. Rev. D \textbf{84}, 066003 (2011)
[arXiv:1108.0400 [hep-th]].

\bibitem{AdSwaves}
E.~Ayon-Beato, G.~Giribet and M.~Hassaine,
``Bending AdS Waves with New Massive Gravity,''
JHEP \textbf{05}, 029 (2009)
[arXiv:0904.0668 [hep-th]].

\bibitem{Donnay}
L.~Donnay, G.~Giribet and J.~Oliva,
``Horizon symmetries and hairy black holes in AdS,''
JHEP \textbf{09}, 120 (2020)
[arXiv:2007.08422 [hep-th]].

\bibitem{FuentealbaHenneaux}
O.~Fuentealba, M.~Henneaux and C.~Troessaert,
``Logarithmic supertranslations and supertranslation-invariant Lorentz charges,''
JHEP \textbf{02}, 248 (2023)
[arXiv:2211.10941 [hep-th]].

\bibitem{Paulos}
M.~F.~Paulos,
``New massive gravity extended with an arbitrary number of curvature corrections,''
Phys. Rev. D \textbf{82}, 084042 (2010)
[arXiv:1005.1646 [hep-th]].

\bibitem{Shina}
A.~Sinha,
``On the new massive gravity and AdS/CFT,''
JHEP \textbf{06}, 061 (2010)
[arXiv:1003.0683 [hep-th]].





\bibitem{Ferreira:2014ina}
H.~R.~C.~Ferreira and J.~Louko,
Phys. Rev. D \textbf{91} (2015) no.2, 024038
doi:10.1103/PhysRevD.91.024038
[arXiv:1410.5983 [gr-qc]].



\end{thebibliography}
\end{document}